\documentclass{article}
\usepackage{graphics}
\usepackage{icrctc07,amssymb,amsmath,times}
\title{A search for possible anisotropies of cosmic rays with $0.1<E<10$ EeV in the region of the Galactic Centre}
\shorttitle{A search for possible anisotropies...}

\authors{E.M Santos$^1$, for the Pierre Auger Collaboration$^2$}

\afiliations{$^1$Centro Brasileiro de Pesquisas F\'{i}sicas, R. Xavier Sigaud, 150, 22290-180, Rio de Janeiro, RJ, Brazil\\
$^2$Observatorio Pierre Auger, Av. San Mart\'{i}n Norte 304, (5613) Malarg\"{u}e, Mendoza, Argentina} 

\email{emoura@cbpf.br}
 
\abstract{
We present updated results for anisotropy searches in the direction of the Galactic Center (GC) at energies 
in the ranges: $0.1<E<1$ EeV and $1<E<10$ EeV. We use data from the Pierre Auger Observatory up to March, 
2007. Present analyzes are therefore based on a substantially larger data set than our previous published 
results. A limit on the flux coming from a hypothetical point-like neutron source at the GC for $1<E<10$ 
EeV was imposed, and searches for extended excesses were also performed.}

\begin{document}
\maketitle

\section[Introduction]{Introduction} 
\label{intro}
\vspace{-0.3cm}
The widely accepted idea that the Galactic Centre harbors a super massive blackhole which would be 
associated to the radio emissions from Sagittarius A*, together with the recent measurements by the 
H.E.S.S. collaboration of TeV  $\gamma$-ray emissions close to the location of Sagittarius A* \cite{HESS}, 
makes the GC a very interesting place to look for localized excesses of cosmic rays at EeV energies 
(1 EeV = $10^{18}$ eV), especially for Auger, where it passes only 6$^\circ$ away from the observatory 
zenith. Also, there have been claims in the past from the AGASA \cite{AGASA} and SUGAR \cite{SUGAR} 
experiments of large excesses in this region, 
none of them were confirmed by a later analysis of Auger data \cite{gcpaper}. Recently, several scenarios 
predicting neutron fluxes from the GC have been discussed in the literature \cite{neutrons}, so that it is 
interesting to update previous Auger bounds with the larger accumulated statistics of the surface array. 
The bulk of cosmic rays consists of charged particles, which are strongly deflected in the galactic magnetic 
field. We therefore perform as well an extended source search around the GC for windows sizes of 10$^\circ$ 
and 20$^\circ$. We divide our data set into 2 energy bands: $0.1<E<1$ EeV and $1<E<10$ EeV. Both ranges 
are studied in terms of Li-Ma overdensities significances.

\section[Data Set]{Data Set}
\label{dataset} 
\begin{figure}[t]
  \begin{center}
    \includegraphics[width=0.42\textwidth]{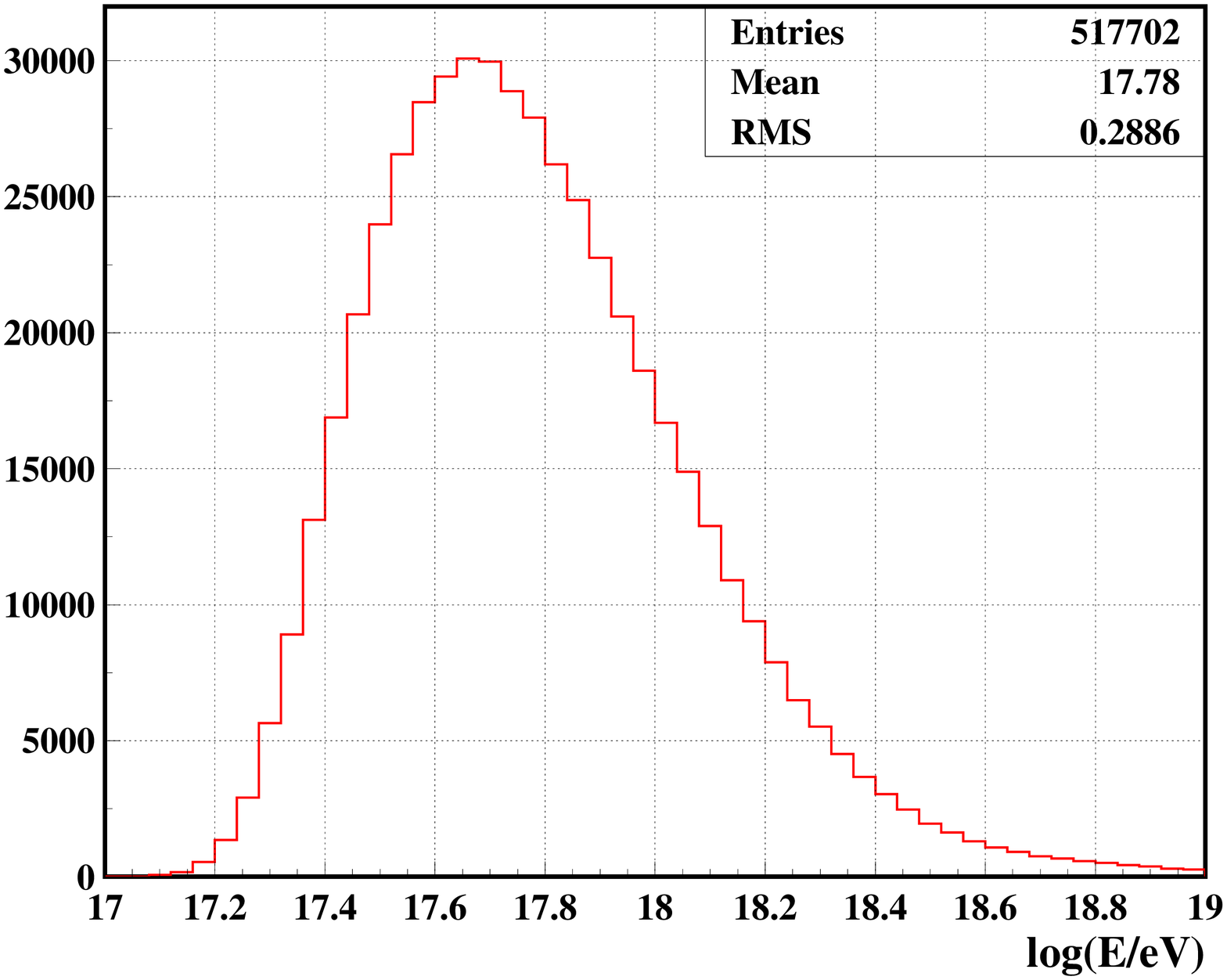}
    \vspace{-0.4cm} 
    \caption{Energy distribution of the events after selection cuts showing the low energy SD threshold.
      \label{ehisto}}
  \end{center} 
\end{figure}

\begin{figure}[t]
  \begin{center}
    \includegraphics[width=0.42\textwidth]{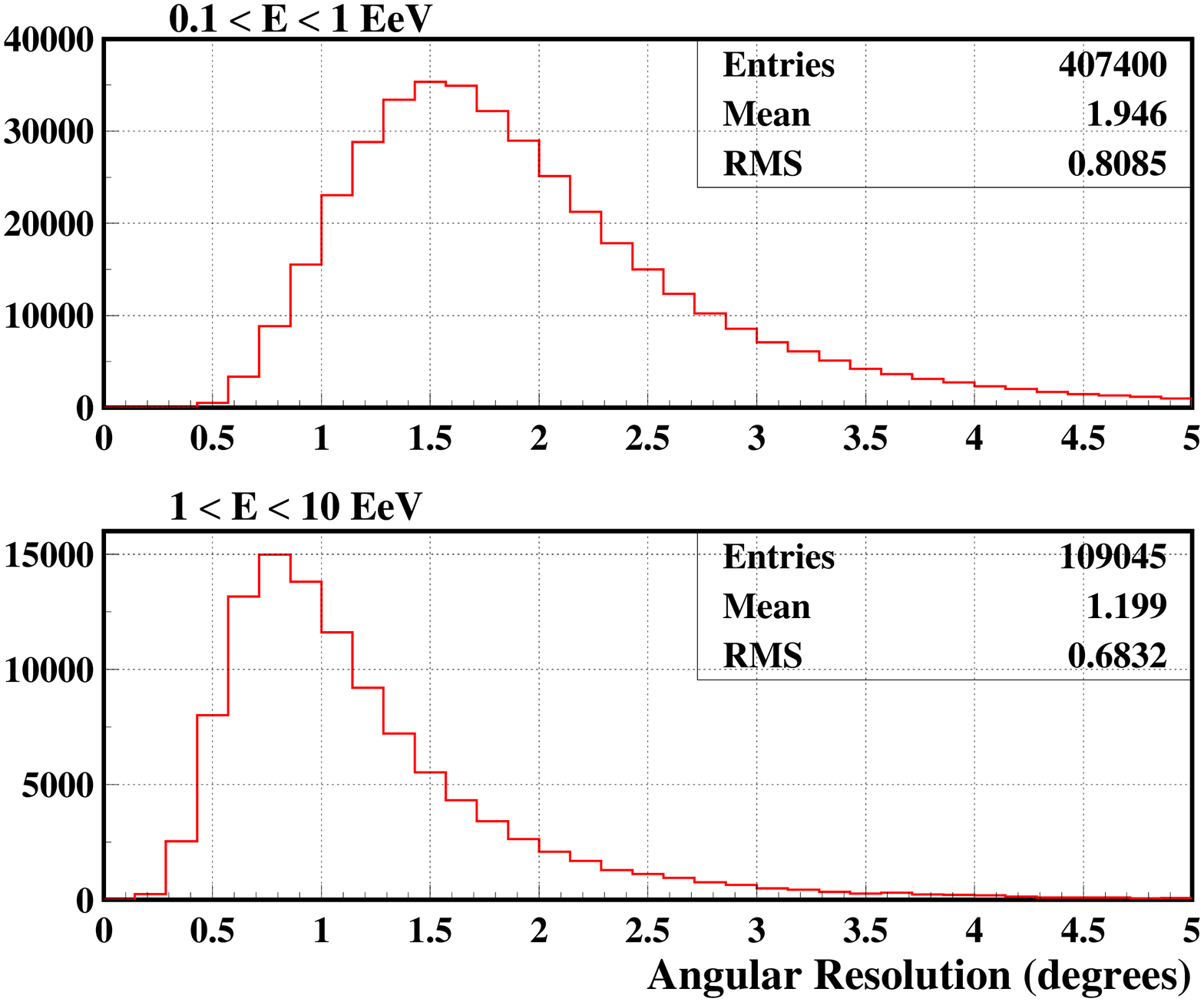}
    \vspace{-0.4cm} 
    \caption{Angular resolution ($1.5\sigma$) distribution for 
    $0.1<E<1$ EeV and $1<E<10$ EeV.\label{angres}}
  \end{center} 
\end{figure}

\vspace{-0.3cm}
For the present analysis, we use data collected by the Pierre Auger Observatory from January 1st, 2004 until 
March 31st, 2007, representing a considerable enhancement in the statistics compared to previous Auger published 
results. We consider surface detector (SD) events where 3 or more tanks in a compact configuration are triggered. 
The events are required to pass the level 5 trigger \cite{icrc2005-trigger}, for which the station with the 
highest signal has to be surrounded by an hexagon of working tanks. Such a requirement ensures a good 
reconstruction of the event. Finally, only showers with zenith angle $\theta<60^\circ$ are used. The energy 
assignment for each shower is done by correlating a SD energy estimator (in this case, a variable related to the 
time integrated water Cerenkov signal at 1000 m from the core) with the calorimetric energy measurement performed by the 
fluorescence detector. This is the procedure used to build the Auger spectrum from SD data and it is 
discussed in details in \cite{icrc2007-spec}. The final data sample energy distribution is plotted in figure 
\ref{ehisto}. For the events used in this analysis, the angular resolution 
\cite{icrc2007-AR} is about 2.0$^\circ$ for $0.1<E<1$ EeV, where the sample is dominated by showers triggering 
only 3 stations and 1.2$^\circ$ for $1<E<10$ EeV, with the enhancement of higher multiplicity events (see mean 
values of the histograms in figure \ref{angres}).

\subsection[Galactic Center]{Searches around the Galactic Centre}
\label{GC}
To impose upper limits on the flux of a hypothetical neutral particles source at the GC, we have done the 
search using a Gaussian filter matching the experiment angular resolution, since this leads to an optimal 
search as shown in \cite{Billoir-ALS}. We have assumed the following differential form for the CR spectrum 
\begin{equation*}
\Phi_{CR}(E) = \kappa 50 \left(\frac{E}{\textrm{EeV}}\right)^{-3.3} \!\!\!\textrm{EeV}^{-1} \textrm{km}^{-2} 
\textrm{yr}^{-1}\textrm{sr}^{-1}\!\!\!, 
\end{equation*}
which has a slope consistent with the HiRes experiment in the energy range $10^{17.5} < E < 10^{18.5}$ eV 
and is a smooth extrapolation of the Auger spectrum measured above 3 EeV. The factor $\kappa$ is introduced 
to take into account the limited knowledge of the true CR flux and should be of order unity for the existing 
measurements at EeV energies ($\kappa=1.2$ if the HiRes normalization is the correct one, around 2 for 
AGASA and 1 for Auger). The 95\% CL upper bound in the integrated (and acceptance weighted) source flux is then given by  
\begin{equation*}
  \int\limits_{E_{min}}^{E_{max}}\!\!\!\!\! dE {\cal A}_{s}\Phi_{s}^{95}(E) = \frac{n_{s}^{95}}{n_{exp}}4\pi\sigma^{2}
  \!\!\!\!\!\int\limits_{E_{min}}^{E_{max}}\!\!\!\!\! dE {\cal A}_{CR}\Phi_{CR}(E),
\end{equation*}
where $n_{s}^{95}$ is the signal counts upper bound at 95\% CL for a given observed/expected ratio and $n_{exp}$ 
is the corresponding background counts inside the Gaussian window of size 
$\sigma$. The background estimation is done using coverage maps, which take into account as much of the local 
effects over the acceptance as possible, like detector deadtimes, array growth and weather effects. Therefore, to 
actually impose the limit, the acceptances ${\cal A}_{s}(E)$ and ${\cal A}_{CR}(E)$ for the source particles and 
for the CR bulk, as a function of energy, must be well understood, especially at low energies, 
where the SD detector is not fully efficient, as can be seen from the strong suppression close to the SD 
threshold (see figure \ref{ehisto}). 

\subsection{$0.1<E<1$ EeV}
Below 1 EeV, neutrons from the GC are strongly depleted by beta decays. An interesting prospect for data 
below 1 EeV would then be the possibility of imposing upper limits on the flux of ultra high energy photons. 
The H.E.S.S collaboration has reported a remarkably flat spectrum of gamma-rays above 165 GeV from 
Sagittarius A*. A naive extrapolation of this spectrum would lead to a flux of photons of 
0.58 km$^{-2}$ yr$^{-1}$ and 0.14 km$^{-2}$ yr$^{-1}$ above $10^{17}$ eV and $10^{17.5}$ eV, respectively. 
In table \ref{ratios1} we show the observed/expected ratio for a point-like search using a Gaussian beam with 
$\sigma=1.3^\circ$ and the corresponding 95\% CL bound on the source counts. No significant excess is seen. 
As discussed above, the conversion of $n_{s}^{95}$ 
into a photon flux upper bound needs a dedicated study of the photon acceptance\footnote{For the case of photons, 
not only we know the acceptance to the bulk (nucleons and heavier nuclei) is different from the source, but 
also the energy assignments.}.

We have done as well a search for extended excesses around the GC on top-hat windows of 10$^\circ$ and 20$^\circ$ 
and the results are presented in table \ref{ratios1}, showing no significant excesses. The isotropy of the sky 
around the GC for $0.1<E<1$ EeV can also be seen from figures \ref{sigmap} C) and D), which are, respectively, 
a map of overdensity significances and the histogram of such significances, together with the expectations of 
isotropy. Given the statistics at each bin, we expect that the Poisson distribution can be fairly approximated 
by a Gaussian, which is plotted in red.
\begin{table*}[t]
  \begin{center}
    \begin{tabular}{clrl}
      \hline
      \hline
      \multicolumn{1}{c}{search} & \multicolumn{1}{l}{window size} & \multicolumn{1}{c}{$n_{obs}/n_{exp}$} & 
      \multicolumn{1}{c}{$n_{s}^{95}$} \\
      \hline
      \hline
      extended    & 10$^\circ$  (TH) &  5663/5657  = 1.00 $\pm$ 0.02(stat) $\pm$ 0.01(syst) &      \\
                  & 20$^\circ$  (TH) & 22274/22440 = 0.99 $\pm$ 0.01(stat) $\pm$ 0.01(syst) &      \\
      \hline
      point-like  & 1.3$^\circ$ (G)  & 192.1/191.2 = 1.00 $\pm$ 0.07(stat) $\pm$ 0.01(syst) & 17.9 \\
      \hline
      \hline
    \end{tabular}
    \caption{Summary of excess searches below 1 EeV around the GC (Sagittarius A*) in the form of 
      extended sources using top-hat (TH) beams and point-like sources with Gaussian (G) beams. A 1\% 
      systematic contribution from possible weather and detector deadtime induced rate modulation 
      is included. The Gauss beam matches the angular resolution (AR=1.5$\sigma=2^\circ$ \cite{icrc2007-AR}) 
      in this energy range. 408607 showers selected.\label{ratios1}}
  \end{center}
  \vspace{-0.3cm}
\end{table*}
\begin{table*}[t]
  \begin{center}
    \begin{tabular}{clrlc}
      \hline
      \hline
      \multicolumn{1}{c}{search} & \multicolumn{1}{l}{window size} & \multicolumn{1}{c}{$n_{obs}/n_{exp}$} & 
      \multicolumn{1}{c}{$n_{s}^{95}$} & \multicolumn{1}{c}{$\Phi_{s}^{95}$} (km$^{-2}$ yr$^{-1}$)  \\
      \hline
      \hline
      extended    & 10$^\circ$  (TH) &  1463/1365  = 1.07 $\pm$ 0.04(stat) $\pm$ 0.01(syst) & &     \\
                  & 20$^\circ$  (TH) &  5559/5407  = 1.03 $\pm$ 0.02(stat) $\pm$ 0.01(syst) & &     \\
      \hline
      point-like  & 0.8$^\circ$ (G)  & 16.9/17.0   = 0.99 $\pm$ 0.17(stat) $\pm$ 0.01(syst) & 5.6 & 0.018$\kappa$ \\
      \hline
      \hline
    \end{tabular}
    \caption{Same as table \ref{ratios1} for $1<E<10$ EeV, including also 95\% CL flux upper limits 
      on a neutron source. The Gauss beam used matches the angular resolution 
      (AR=1.5$\sigma=1.2^\circ$ \cite{icrc2007-AR}) in this energy range. 109101 showers selected.
      \label{ratios2}}
  \end{center}
  \vspace{-0.5cm}
\end{table*}

\subsection{$1<E<10$ EeV} 
\begin{figure*}[t]
  \begin{minipage}{\textwidth} 
    \begin{minipage}[t]{0.5\textwidth} 
      \begin{center}
	\includegraphics[width=0.83\textwidth]{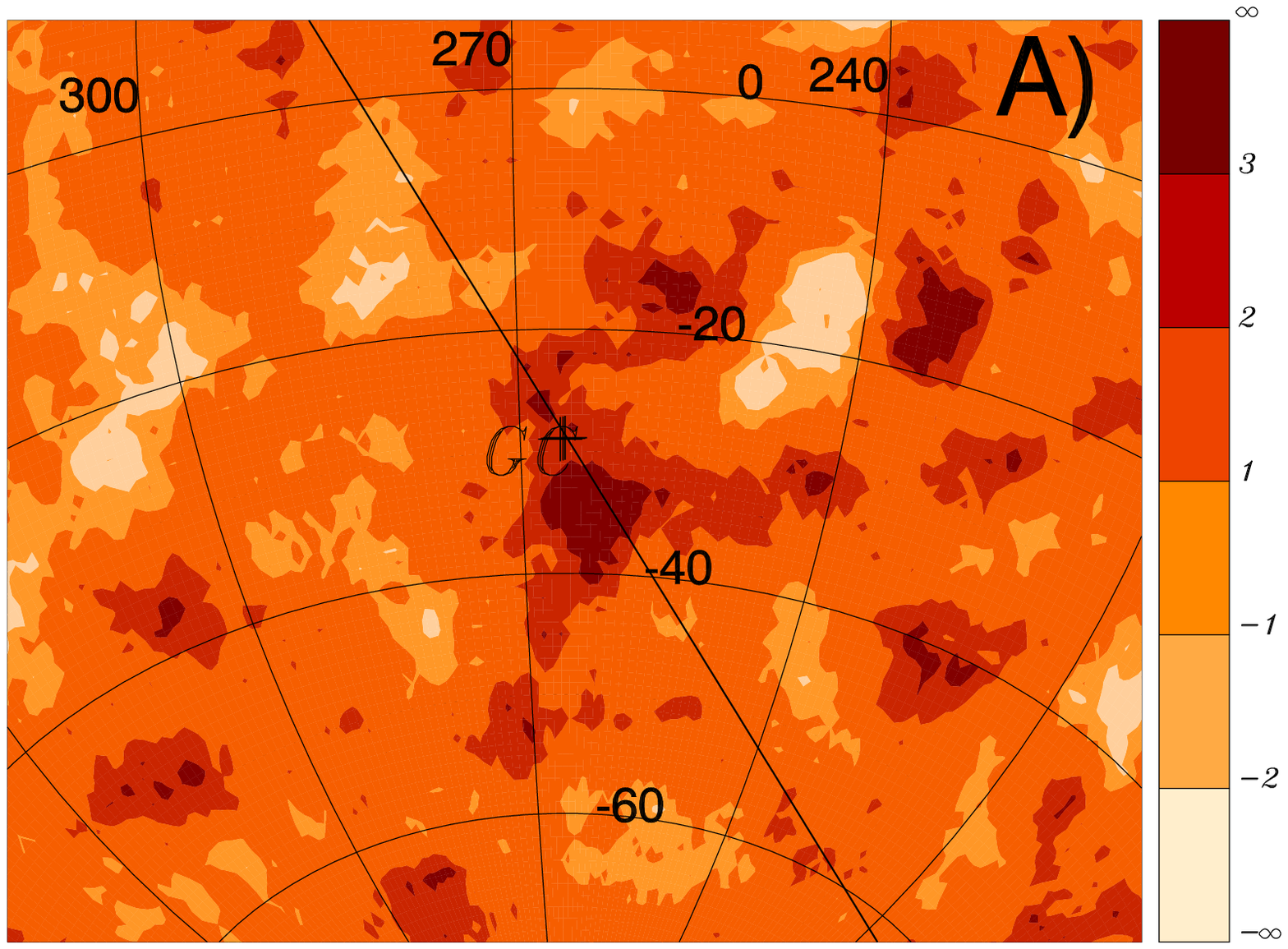}
      \end{center} 
    \end{minipage} 
    \begin{minipage}[t]{0.5\textwidth} 
      \begin{center} 
	\vspace{-4.8cm} 
	\includegraphics[width=0.8\textwidth,height=0.7\textwidth]{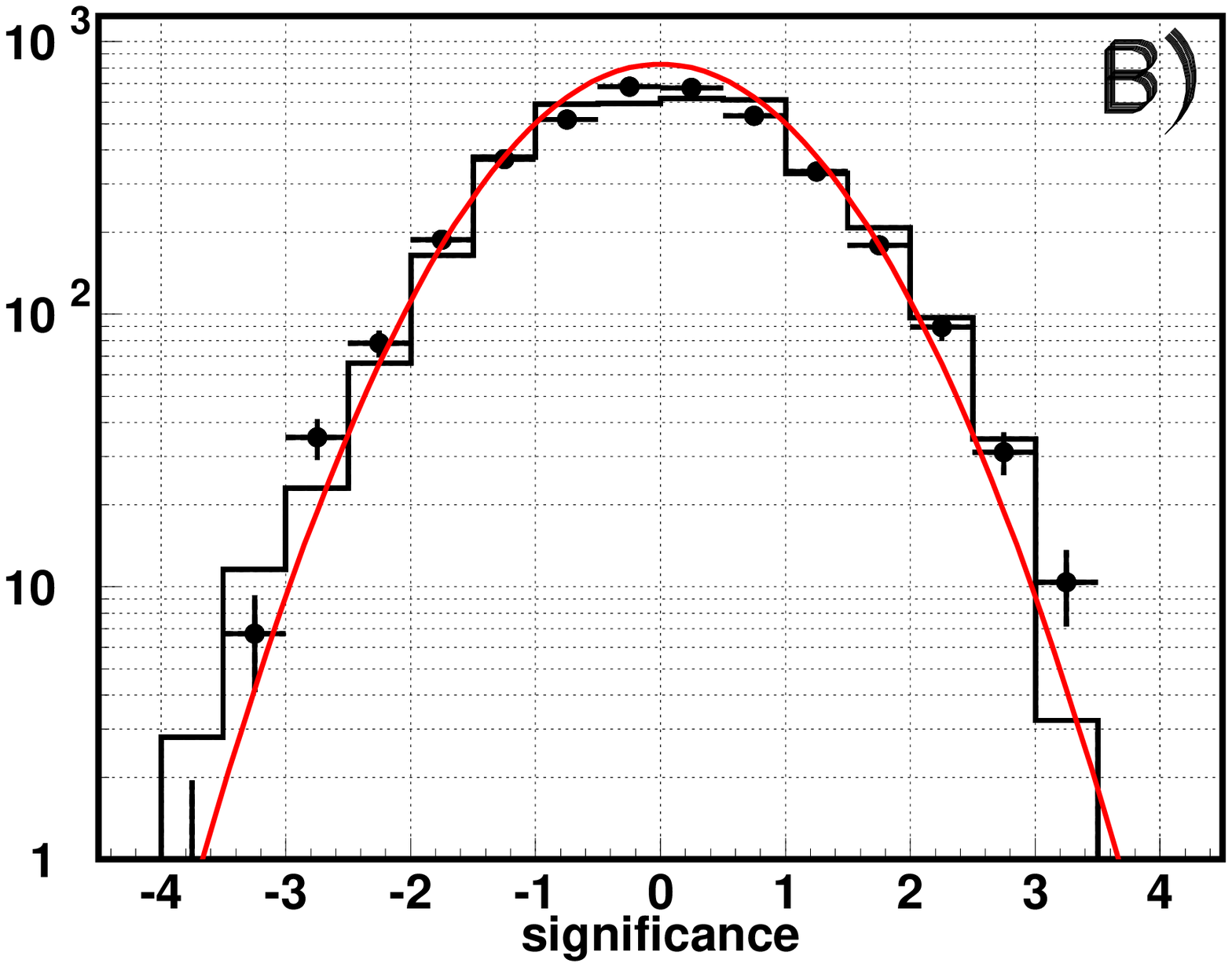}
	\vspace{-0.2cm} 
      \end{center} 
    \end{minipage} 
    \begin{minipage}[t]{0.5\textwidth} 
      \begin{center}
	\includegraphics[width=0.83\textwidth]{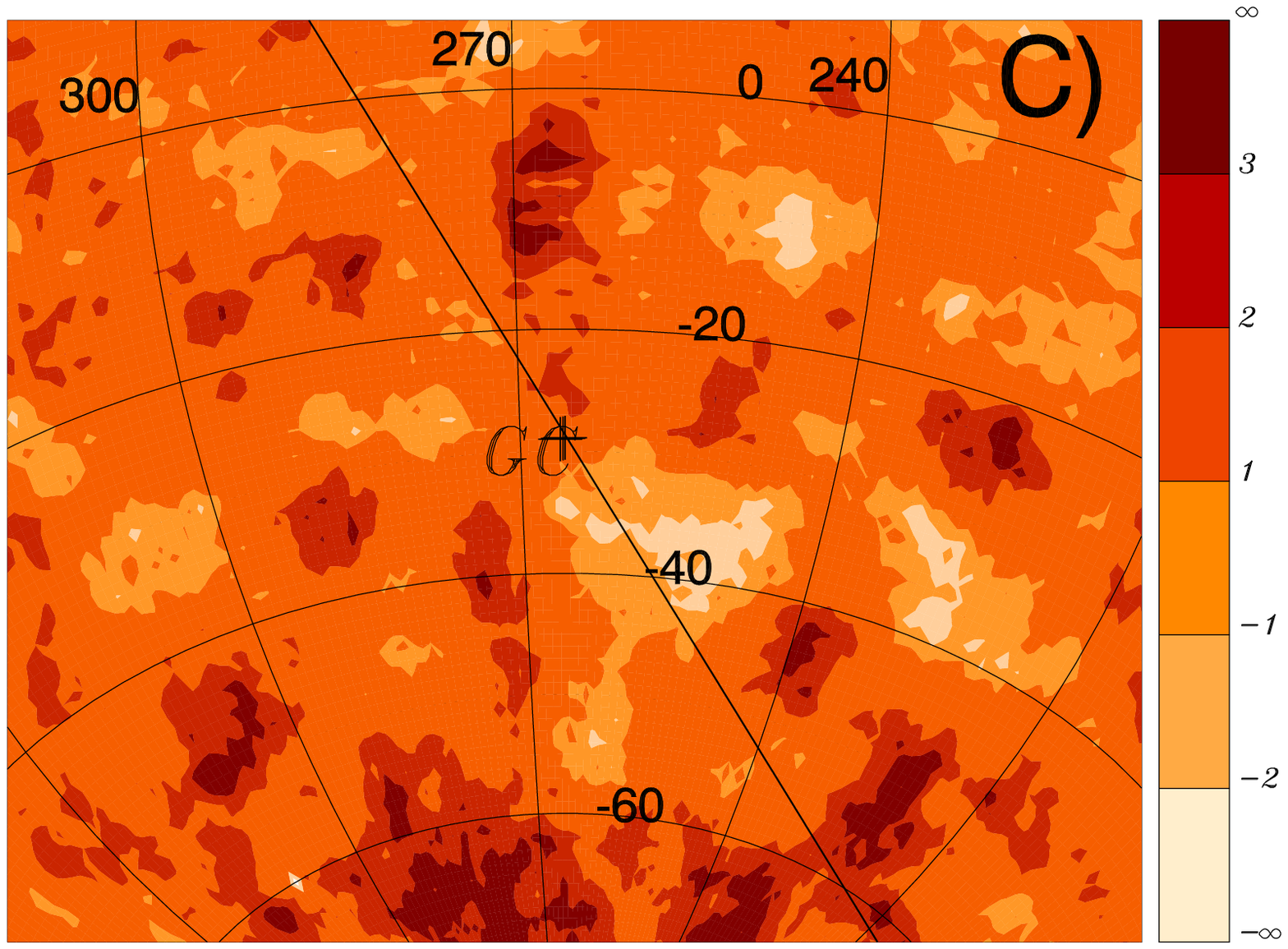}
      \end{center} 
    \end{minipage} 
    \begin{minipage}[t]{0.5\textwidth} 
      \begin{center} 
	\vspace{-4.8cm} 
	\includegraphics[width=0.8\textwidth,height=0.7\textwidth]{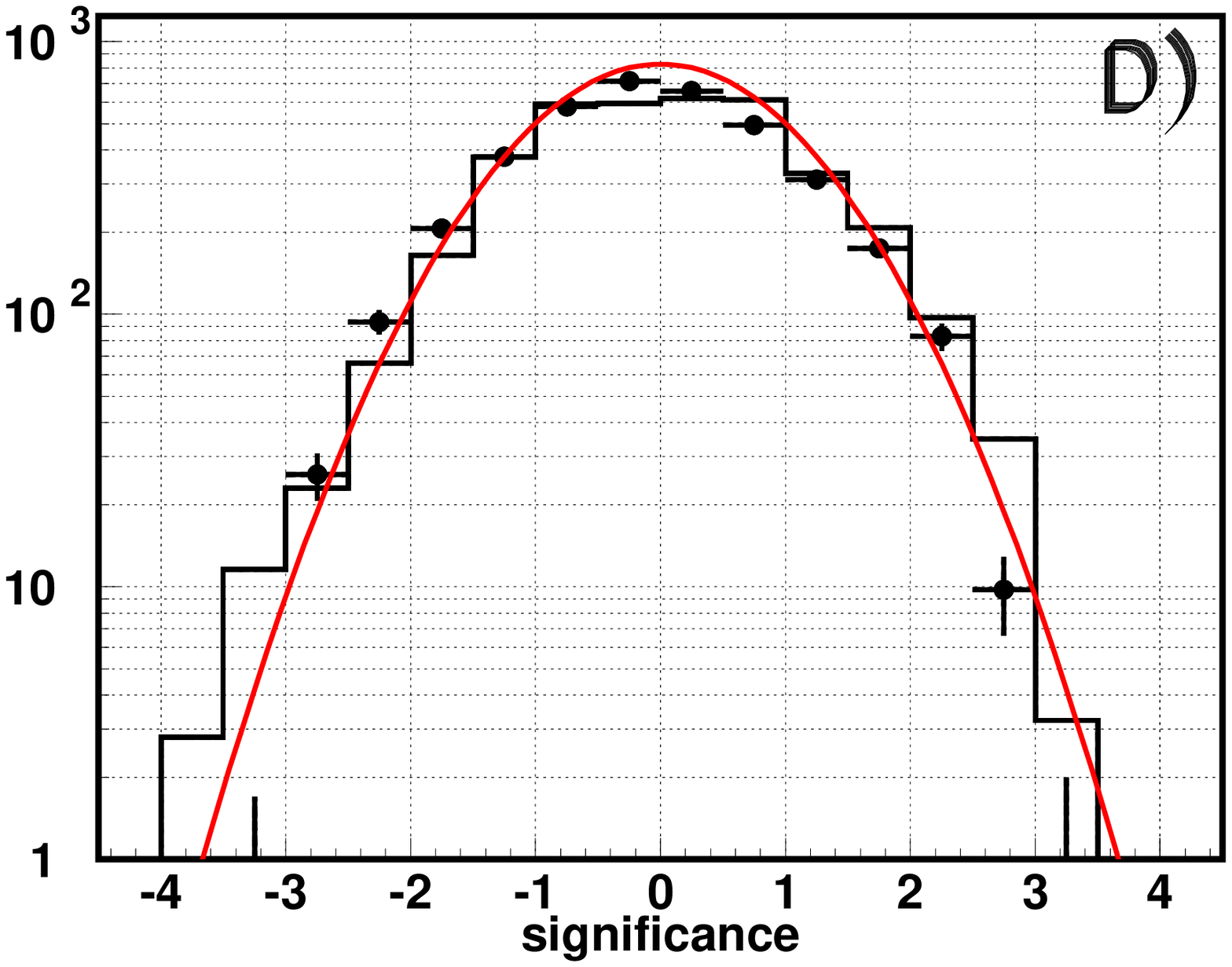}
	\vspace{-0.2cm} 
      \end{center} 
    \end{minipage} 
    \caption{A) and C): Maps of the CR overdensity significances near the GC region on top-hat windows of 5$^\circ$ 
      radius for $1 < E < 10$ EeV and $0.1<E<1$ EeV, respectively. The GC location is indicated with a cross, lying along 
      the galactic plane (solid line). B) and D): Histograms of overdensity significances for the same energy ranges. The 
      dots represent the data, the solid histogram is the MC prediction of an isotropic sky and the expected Gaussian 
      fluctuations are given by the red solid curve.
      \label{sigmap}}
  \end{minipage}  
\end{figure*} 

Given the neutron lifetime at rest (885.7s), a considerable fraction of neutrons produced in the $1 < E < 10$ EeV 
energy range at the GC would survive and reach the Earth. Even though energies above 3~EeV are probably too high 
to be reached by particles accelerated by galactic objects, in this energy range, the bulk of CRs will be below 
that value. Unlikely the case of photons, if we assume here that the composition of the bulk of the CRs in the 
energy range studied is similar to the source, so that the source spectrum $\Phi_{s}$ has the same spectral index as 
the CR spectrum $\Phi_{CR}$, we can obtain directly the 95\% CL upper bound in the integrated flux (see 
numbers in table \ref{ratios2} where we also show the results of extended source counting) as
\begin{equation*}
  \Phi_{s}^{95} = \frac{n_{s}^{95}}{n_{exp}}4\pi\sigma^{2}\int\limits_{E_{min}}^{E_{max}} dE \Phi_{CR}(E),
\end{equation*}
which can be finally converted into an upper limit on the source luminosity of 
${\cal L}_{GC}<1.25\kappa \times 10^{34}$ erg/s. Such a limit already excludes most of the models for neutron 
production at the GC \cite{neutrons}.

In figure \ref{sigmap} A) one shows the Li-Ma \cite{Li-Ma} significance distribution of overdensities in 5$^\circ$ top-hat 
windows in the region around the GC for the energy range $1<E<10$ EeV. The distribution of the significances of figure 
\ref{sigmap} A) is shown in figure \ref{sigmap} B), together with the expectations from an isotropic sky. The overdensities 
are consistent with an isotropic sky.

\vspace{-0.3cm}
\section{Conclusions} 
\label{concl}
\vspace{-0.3cm}
With an exposure two times larger than the one used in the previous Auger published result, we have found no 
significant CRs flux excess in the region around the Galactic Center in 2 different energy ranges: $0.1<E<1$ EeV 
and $1 < E < 10$ EeV. We have treated the GC both as an extended and a point-like source. The distribution of Li-Ma 
overdensity significances in this region shows consistency with an isotropic sky for both energy ranges. For 
$1<E<10$ EeV, an upper limit on the flux of a hypothetical point-like neutron source in the direction of 
Sagittarius A* was imposed, which excludes most of the models predicting neutrons from the GC.

\end{document}